\documentclass[12pt,usletter]{article} 

\usepackage{charter} 

\usepackage{cite}
\usepackage{graphicx} 
\usepackage{amssymb,amsmath} 
\usepackage{multicol} 
\usepackage{url} 
\usepackage{enumitem} 
\usepackage{marvosym} 
\usepackage{wrapfig} 
\usepackage[T1]{fontenc} 
\usepackage{datetime} 
\newdateformat{mydate}{\monthname[\THEMONTH] \THEYEAR} 
\usepackage[hidelinks]{hyperref}
\usepackage{fancyhdr} 
\usepackage[object=vectorian]{pgfornament} 
\usepackage{tikz}

\usepackage{wasysym}

\newcommand{\sectionlinetwo}[2]{%
  \nointerlineskip \vspace{.5\baselineskip}\hspace{\fill}
  {\color{#1}
    \resizebox{0.5\linewidth}{2ex}
    {{%
    {\begin{tikzpicture}
    \node  (C) at (0,0) {};
    \node (D) at (8,0) {};
    \path (C) to [ornament=88] (D);
    \end{tikzpicture}}}}}%
    \hspace{\fill}
    \par\nointerlineskip \vspace{.5\baselineskip}
  }

\newcommand{\ReportName}[1]{ 
\begin{center}
\Large \usefont{T1}{fvs}{b}{n} 
#1
\end{center}	
\par \normalsize \normalfont}

\begin{document}

\ReportName{Communicating Intel to Decision-Makers:\\ Toward the Integration Text and Charts in Reports} 
\begin{center}
    {\large Melanie Bancilhon, R. Jordan Crouser, Alvitta Ottley}
\end{center}
{\centering}


\vspace{0.5cm}

\sectionlinetwo{darkgray}{88}

\begin{center}
\begin{minipage}[h]{0.75\linewidth}

Intelligence analysts' roles include analyzing reports, identifying important information relevant to the current state of affairs, and communicating their takeaways. These reports are then analyzed and reported to decision-makers or translated into a presidential brief. While these tasks seem consistent across analysts, each step differs in its content, level of detail, and format. The purpose of this research is to gain an understanding of how consumers of analytic products receive and interact with reports. This was accomplished via a series of online questions to 22 experts recruited to provide input. Our analysis provides insight into what makes analytic products effective for decision-makers, which could improve the quality of reports produced and alleviate common customer pain points, should recommendations be appropriately incorporated.

\end{minipage}
\end{center}

\vspace{0.5cm}

\setlength{\columnsep}{40pt} 
\begin{multicols}{2} 


\section*{Introduction}
The primary objective of intelligence analysis is to deliver ``the right information, to the right person, at the right time, communicated in the right way.'' The outputs of such analysis, or \emph{analytic products}, may take the form of intermediary data (e.g., a curated or enriched dataset) intended to be used as input for subsequent analysis or end-products (e.g., written reports, interactive dashboards, daily briefings, and placemats). These aim to provide decision-makers with a concise, well-supported discussion of the findings and associated recommendations.

Digital intelligence reports allow various formatting options and include images and links (e.g., to other reports, web pages, or audio and video files). These reports also allow for user interactions such as highlighting and note-taking. Little is known about how report formats vary across tasks and analytical workflows and how analysts interact with different report formats. However, we can develop better tools tailored to different analytical workflows and tasks by gaining insight into analytic report formats and user interactions. 

At present, the personal tailoring of analytic end-products is a painstakingly-manual process. Because of its labor-intensive nature, it is often only those in positions of substantial power who can reap the benefits of bespoke analytic products. However, the quality of decisions made ``at the top'' is fundamentally bounded by the quality of decisions made at all preceding levels. Moreover,  prior studies suggest that even when presented with all of the relevant information, decision quality suffers when presented in a format poorly suited to the individual decision-maker.

We share the results of 22 interviews with decisions makers in the intelligence community to understand the existing practices for information dissemination and how decision-makers ingest analytic end-products. Using this information, we propose design guidelines for developing reporting tools that can readily adapt to the needs of individual decision-makers.

\section*{Background}

The effect of adding images to text has been studied for many decades in the Psychology literature ~\cite{levie1982effects}. Researchers argue that illustrations contextualize information contained in the text and help the viewer to organize and interpret it~\cite{bransford1972context}. For example, studies suggest that adding relevant images to text improves comprehension and retention~\cite{holiday1967stem}. One study has shown that the textual component can be the most memorable part of a visualization in a memory task \cite{borkin2015beyond}. Bruny\'{e} et al. demonstrated the value of multimedia formats for improving performance: participants who relied on instructions combining text and diagrams assembled a toy significantly faster and more accurately than those given text-only or illustration-only instructions~\cite{brunye2006learning}. 

Although the exact cognitive process underlying such betterment remains debated ~\cite{levie1982effects}, the benefits of pairing text with relevant graphics are particularly interesting to the Visualization community. In 2013, Kosara et al.~\cite{kosara2013storytelling} proclaimed \textbf{Narrative/Storytelling Visualizations} (combing text and visualization) to be ``the next step for visualization'', and researchers have explored different storytelling techniques~\cite{hullman2011visualization, hullman2013deeper,segel2010narrative}.

There has been extensive research in multimedia and data visualization that shows that data format can impact perception, reasoning, and decision-making\cite{micallef2012assessing,ottley2012visually,ottley2015improving,ottley2019curious,mosca2021does,bancilhon2023combining,zhi2019linking}. Most relevant to the current work, Zhi et al. \cite{zhi2019linking} have examined how different ways of integrating text and visualization impact comprehension and engagement in a storytelling scenario. They found that comprehension scores were higher amongst all formats with slideshow layouts, where each frame contains the text and corresponding visualization displayed side by side and represents an event in the story. Additionally, Stokes et al.\cite{stokes2022striking} found that users prefer charts with the largest textual annotations compared to charts with fewer annotations. They also found that more takeaways and relationships are drawn when the text describes statistical or relational components of a chart. Many other researchers have studied variations in personal preferences for charts versus textual representations. 



\begin{figure*}
    \begin{center}
          \includegraphics[width=.8\textwidth]{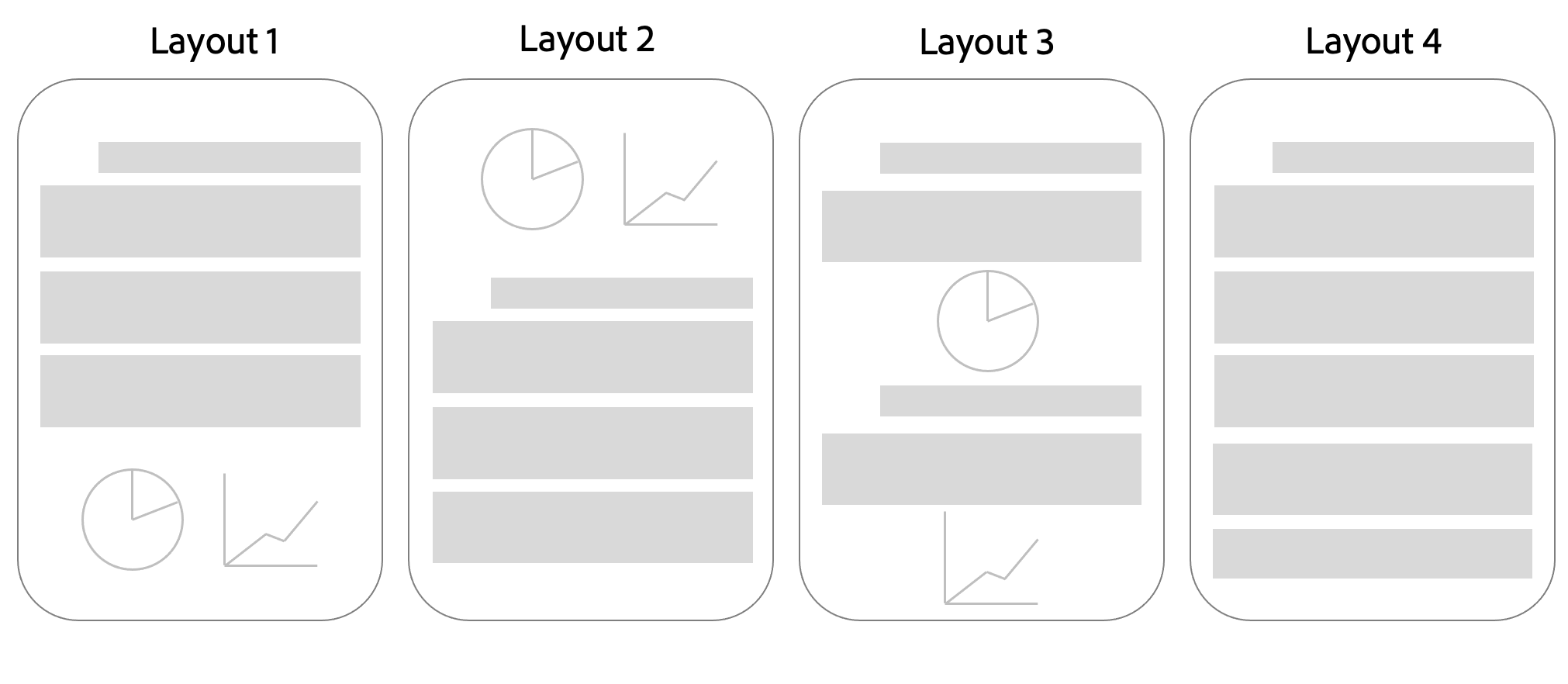}
          \caption{\textbf{Layout 1:} Text followed by charts. \textbf{Layout 2:} Charts followed by text. \textbf{Layout 3:} Charts within text. \textbf{Layout 4:} Text only}
          \label{fig:layouts}
    \end{center}
\end{figure*}

\section*{Interviews with Analysts}
We interviewed 22 professionals working in the intelligence space to gain insights about how they interact with intelligence reports, their most used and preferred report formats, and the types of decisions they make using the information from these reports. The interview contained 18 questions (see appendix material for questions).

We presented analysts with four different potential report layouts. In layout 1, the graphs are presented at the end of the documents. In layout 2, the charts are presented first, followed by text. In layout 3, the charts are presented throughout the report next to the relevant text. Layout 3 is a report format without charts (see figure \ref{fig:layouts}. 

\section*{What Makes a Report Useful?}

The most common responses for what makes a report useful were \textit{Clear and Accurate (n=7)}, \textit{Concise(n=8)}, and \textit{Relevant (n=5)}. For example, one analyst said:

\begin{quotation} 
\noindent{\huge``}

\noindent\normalsize\textit{Report is useful when it clearly identifies and defines concepts and provides understanding into the locations, the importance of the group/person/asset, insight into the need/value of the collection, and has relevant accompanying graphics to amplify data.}

\hfill{\huge''}
\end{quotation}
Analysts reported the following useful features: \textit{tear line}~\footnote{The ``tear line'' is a physical line on a document, often used to separate classified and unclassified information.}, charts and graphics, well-sourced end-notes, easy to transfer and download, summary section, and takeaway section.

\section*{Analysts Reading Patterns}
In order to better understand the layout and report format preferences, we examined analysts' reading patterns. 7 participants stated that they "Quickly scan and jump to the most important content", 5 stated that they "Read the entire report from start to finish" and the rest reported using a combination of techniques, sometimes including both of these extremes. Among the 9 participants that used several techniques, 6 reported that they read all titles and headers first and two reported that they read the summary first.

\section*{The Prevalence of Charts in Intelligence Reports}
We were interested in how often intelligence reports contain charts and graphics. Only one analyst reported that their reports never contain graphics. 8 participants reported that their reports \textit{sometimes} contain graphics, 6 said that they contain graphics \textit{about half of the time}, 6 said that they contain reports \textit{most of the time} and 1 said that they \textit{always} contain graphics.

\begin{figure*}
    \begin{center}
          \includegraphics[width=.8\textwidth]{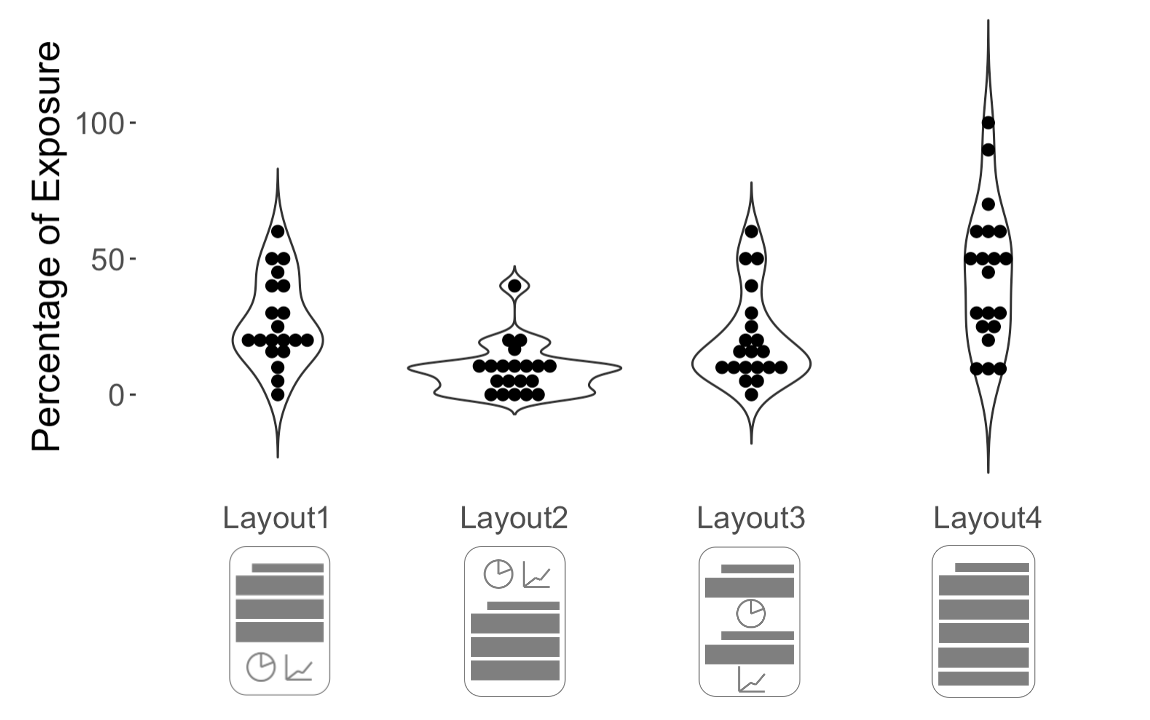}
          \caption{Layout 4 was the most used on average although variability was high, and Layout 2 was the least used.}
           \label{fig:percentageExposure}
      \end{center}
\end{figure*}

\subsection*{Prevalence of Different Layouts in Reports}

We asked analysts to report the percentage of reports that they receive that are presented in similar formats to Layouts 1-4 (e.g. Layout 1 = 70\%, Layout 2 = 30\%). One participant's data was missing so we analyzed the responses of 20 analysts. 

Figure \ref{fig:percentageExposure} shows the prevalence of each layout across all 20 analysts, where each dot represents one participant's response. We can see that roughly half of the participants receive Layout 4 at least 50\% of the time and all of them receive Layout 2 at most 50\% of the time. We computed the average of the reported percentages across each layout. We found that overall, there is a 4.5\% chance that a random analyst receives Layout 1, 0.5\% that they receive Layout 2, 6\% chance that they receive Layout 3, and 15.25\% chance that they receive Layout 4. To understand the use of different report formats with more granularity, we compared exposure to types of decisions to uncover any patterns in the next section.  


\begin{figure*}
    \begin{center}
          \includegraphics[width=1.0\textwidth]{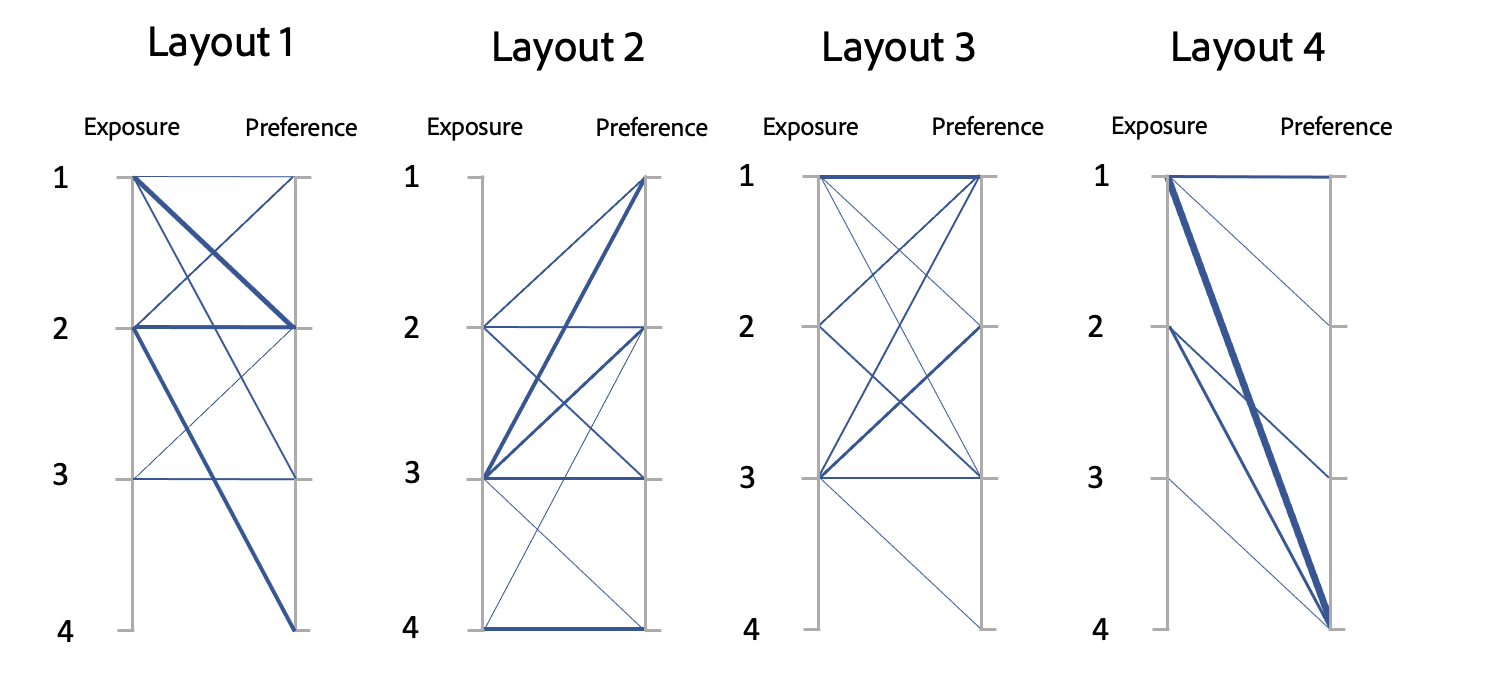}
          \caption{We compared analysts' rankings of most to least consumed report formats to most to least preferred report formats. The width of the lines corresponds to the number of analysts with the corresponding ranking ratios.}
           \label{fig:exposure_vs_preference}
      \end{center}
\end{figure*}

\subsection*{Do Analysts Receive Reports in Their Preferred Layouts?}

We asked analysts to rank their preferred layouts and justify their rankings. Out of the responses from 19 analysts, Layout 3 was ranked first the most (n=8), and Layout 4 was ranked last the most (n=14). We examined the rationale for these rankings and found that the analysts who ranked Layout 3 as the most preferred liked to see charts inserted next to their corresponding text to improve flow and comprehension.

\begin{quotation} 
\noindent{\huge``}

\noindent\normalsize\textit{I prefer the graphics to be spread out throughout the text, like in Layout 3, and that they relate to the information within the paragraphs around the graphic. }

\hfill{\huge''}
\end{quotation}

Analysts who ranked Layout 4 as their least preferred layout believe that reports with charts are faster to read, improve information retention and comprehension, and consolidate information in the text. It is important to note that all three analysts who ranked Layout 4 as the most preferred layout deem charts unnecessary for their tasks. 

\begin{quotation} 
\noindent{\huge``}

\noindent\normalsize\textit{Layout 4 is just a wall of words and can be difficult to read without a graphic. I know some reports don't need or come with graphics, but they can be difficult to read if it's a dull topic or too detailed.}

\hfill{\huge''}
\end{quotation}

To compare report layout consumption to preference, we converted participants' percentage of exposure to each layout to rankings. For example, if someone reported being exposed 65\% of the time to Layout 1, 20\% to Layout 2, 10\% to Layout 3, and 5\% to Layout 4, the most to least consumed formats are Layout 1, Layout 2, Layout 3, and Layout 4 in that order. If someone reported being exposed equally to more than one layout, we placed the layouts in the same ranked position. Figure \ref{fig:exposure_vs_preference} shows how exposure rankings compare to preference rankings for each layout. Overall, the most prevalent layouts are also the most preferred for about 42\% of the analysts (n=8). On the other hand, 8 analysts ranked their most prevalent layout as least preferred. We can see that most participants who tend to be most exposed to Layout 4 ranked it as least preferred.

\begin{figure*}
    \begin{center}
          \includegraphics[width=.8\textwidth]{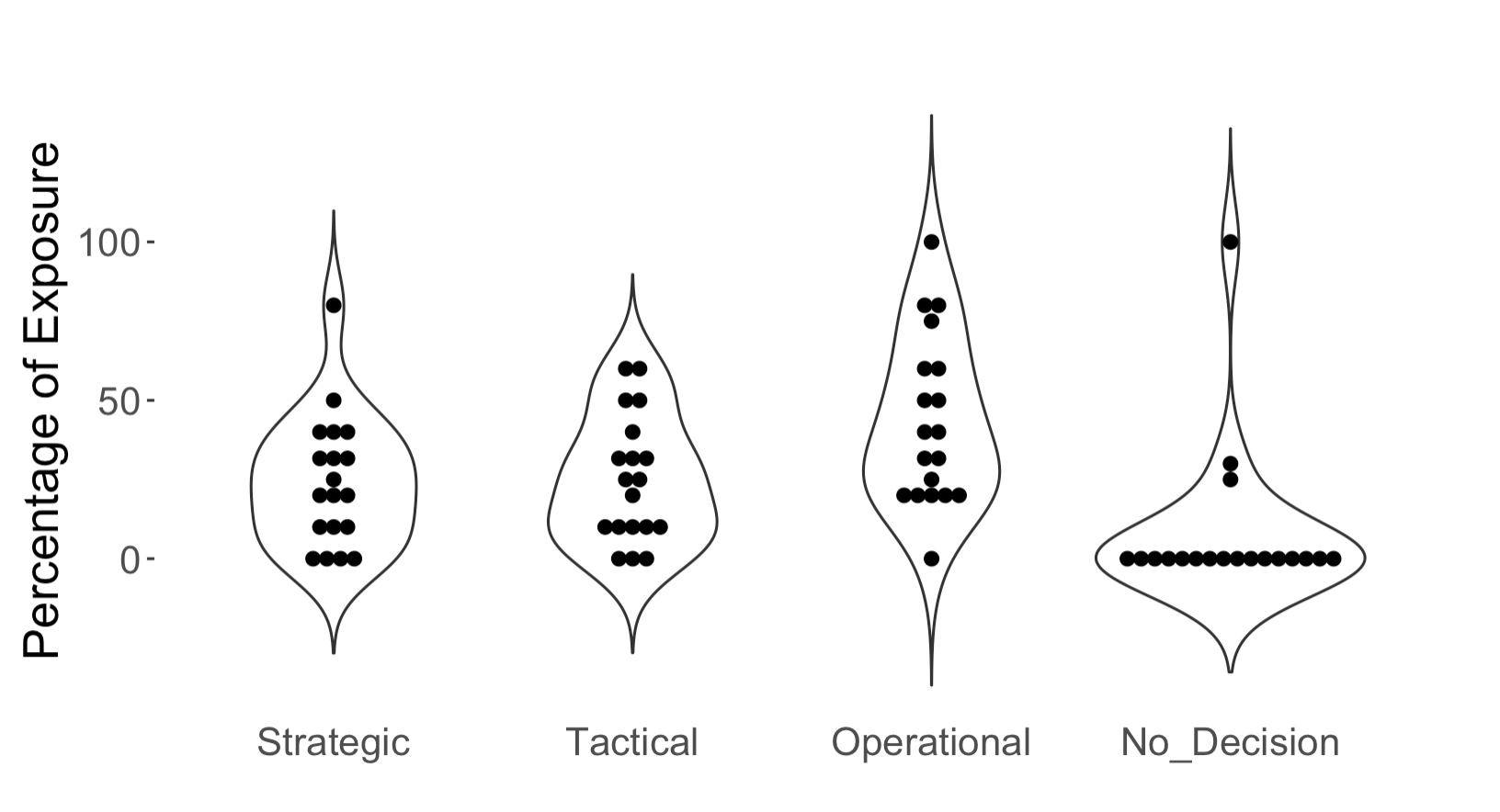}
          \caption{Although differences are small, operational decisions are the most frequent, followed by tactical and strategic, respectively. Three analysts reported conducting tasks not involving decisions. }
           \label{fig:decisions}
      \end{center}
\end{figure*}

\section*{Report Formats and Decision Strategies}

Decisions often differ in nature and require different decision-making processes and decision aids. While there are several ways to classify decision strategies, we adopted the classes of decisions from the business intelligence literature, which define three main types of decisions, namely \textbf{strategic}, \textbf{tactical}, and \textbf{operational}~\cite{schmidt2000strategic,bilgen2004strategic}. 

A strategic decision can be defined as a long-term, complex decision that often has a high impact. They are non-repetitive in nature and involve careful evaluation of several alternatives. Tactical decisions occur with greater frequency as often relate to implementing strategic decisions. Finally, operational decisions relate to daily operations and are often low impact. 

\subsection*{From Report to Decision}
We asked analysts to describe their steps, from reading a report to deciding. We annotated common themes across responses and found that 7 responses included correlating the report with the current effort and identifying operational actions in the report.  

\begin{quotation} 
\noindent{\huge``}

\noindent\normalsize\textit{I must recognize something in the text that can be operationalized or that relates to a specific customer's needs.}

\hfill{\huge''}
\end{quotation}

Another common process mentioned by 6 analysts is cross-referencing with other reports and sources. One analyst mentioned that they verify questionable information, and another makes sure to verify the report's author. \\

\begin{quotation} 
\noindent{\huge``}

\noindent\normalsize\textit{Reading the report, source descriptor, source confidence as well and other reports relevant to a topic all play a role in making a decision. Typically understanding the information, then making sure to put it in its context and reliability, and corroborating information from other sources of intel.}

\hfill{\huge''}
\end{quotation}

\subsection*{Prevalence of Decision Strategies}

Figure \ref{fig:decisions} shows our analysts' prevalence of decision strategies. Three analysts reported that their job also consists of tasks that do not involve deciding. We can observe that the distribution of strategic, tactical, and operational decisions is somewhat even. On average, operational decisions were more frequent, with a mean prevalence of 43.33\%, compared to 24.91 for tactical decisions, 24.12\% for strategic decisions, and 8.16 for tasks not involving decisions. 

\subsection*{Preferred Layout For Each Decision Strategy}

We asked analysts to rank which layout they would prefer for each type of decision. For strategic decisions, layout 3 was preferred (n=7), followed closely by Layout 1 (n=6) and Layout 2 (n=6). The preferred layout for operational decisions was Layout 3 (n=12). The preferred Layout for Tactical decisions was Layout 1 (n=7) and Layout 3 (n=7). 

\begin{figure}[h!]
    \begin{center}
          \includegraphics[width=.48\textwidth]{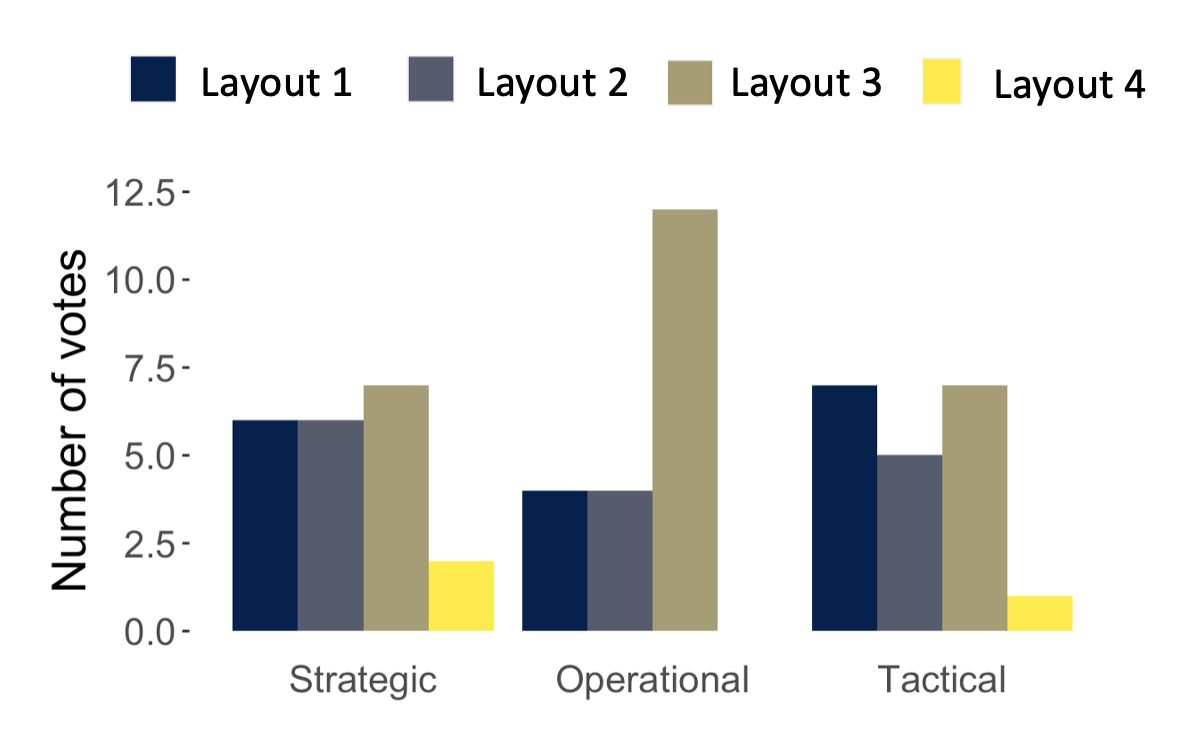}
          \caption{Figure 5: Most ideal layout for each type of decision.}
           \label{fig:decisions_layouts}
      \end{center}
\end{figure}

\section*{Discussion \& Future Work}

Our user study provided valuable insights into how intelligence analysts consume, interact with, and make decisions from reports. Most responses had high variance due to the diverse nature of roles and functions in our pool of analysts. However, we were able to extract some converging themes and patterns in the data.

First, we found that most analysts value clarity, accuracy, conciseness, and relevance the most in reports. Despite this convergence towards these features, analysts reading patterns vary greatly, ranging anywhere between a quick scan and a thorough read. This finding is not surprising as it is likely heavily dependent on the nature of the task. When asking analysts about the steps they take from reading a report to making a decision, several highlighted the importance of linking the information in the report to the situation at hand. Another common procedure several analysts mentioned is cross-referencing the report with other sources. This points to the importance of trust in the report, which includes trust in the accuracy of the information or the report's author. Analysts have to ensure that the information is valid before proceeding. 

We found that, on average, most analysts encounter reports with no charts more frequently than reports with charts. Most of those most exposed to text formats reported preferring their reports to contain charts where applicable. Preference for the juxtaposition of charts and text varies across analysts. Some prefer to have charted within the text for cohesion, some prefer to read the charts first to get an overview, and others prefer to read the text first to provide context to the charts. Analysts also voted for the best layout for each type of decision. We found that for operational decisions, having the charts within the text received more votes. This is likely due to the sequential and straightforward nature of operational decisions. 

Based on our findings, we believe that a personalized interface that accounts for users' reading patterns and preferences, as well as report type and decision type, would benefit intelligence analysts. Further research must be conducted to define the interface's compatibility with the existing workflow. Other more straightforward solutions to improve productivity would be to improve cross-referencing by seamlessly linking similar reports. Moreover, further investigations into the factors that impact report credibility and trust would be beneficial to support implementing systems that detect untrustworthy signals to improve decision quality.

\section*{Acknowledgement}
The authors thank our NCSU collaborators for their expertise and guidance and the 22 decision-makers who participated in this study for their time. This material is based upon work supported in whole or in part with funding from the Laboratory for Analytic Sciences (LAS). Any opinions, findings, conclusions, or recommendations expressed in this material are those of the author(s) and do not necessarily reflect the views of the LAS and/or any agency or entity of the United States Government.

\section*{Appendix: Questionnaire}

\textbf{Section 1 of 2: Report Format and Preferences} \\

\noindent
Reports can come in different formats (e.g., verbal, written, video, and audio). This survey focuses on written documents. 

\begin{enumerate}
    \item How often do interact with written reports?
    \begin{itemize}
        \item [$\ocircle$] Daily   
        \item [$\ocircle$] A few times a week
        \item [$\ocircle$] Weekly
        \item [$\ocircle$] Bi-weekly
        \item [$\ocircle$] Monthly
        \item [$\ocircle$] Other (Please specify) 
    \end{itemize} 

    \item In what format(s) do you usually receive written reports?
    \begin{itemize}
        \item [$\ocircle$] Printed
        \item [$\ocircle$] Digital 
    \end{itemize}

    \item How often do the reports you read contain charts or graphical data representations?
    \begin{itemize}
        \item [$\ocircle$] Always   
        \item [$\ocircle$] Most of the time  
        \item [$\ocircle$] About half the time  
        \item [$\ocircle$] Sometimes
        \item [$\ocircle$] Never
    \end{itemize} 

    \item Do you actively interact with reports by making annotations, notes, or highlighting?
    \begin{itemize}
        \item [$\ocircle$] Yes
        \item [$\ocircle$] No 
    \end{itemize}

    \item How do you read reports?
    \begin{itemize}
        \item [$\ocircle$] Quickly scan and jump to the most important content
        \item [$\ocircle$] Read all titles and headers first 
        \item [$\ocircle$] Read the entire report from end to finish
        \item [$\ocircle$] Look at graphs and pictures first
        \item [$\ocircle$] Other (Please specify) 
    \end{itemize}

    \item What makes a useful report, in your opinion? \\ \\
    \noindent\rule{7cm}{0.4pt}\\ 
    \noindent\rule{7cm}{0.4pt} \\

    \begin{center}
    \hspace{-2em} \textit{\textbf{Below are examples of layouts that you might see in a report.\\}} 
    \includegraphics[width=\linewidth]{figures/visConditions.png}
    \end{center}

    \item Please enter the percentage of reports you currently receive that include these layouts. An example response would be: Layout 1 = 70\%, Layout 2 = 30\%.

    \begin{enumerate}
        \item [i]Layout 1: Text followed by chart(s) \rule{2cm}{0.4pt}
        \item [ii]Layout 2: Chart(s) followed by text \rule{2cm}{0.4pt} 
        \item [iii]Layout 3: Chart(s) within the text \rule{2cm}{0.4pt}
        \item [iv]Layout 4: Text only \rule{2cm}{0.4pt}
    \end{enumerate}

    \item Please refer to the layout images above and rank them in the order that you prefer them. Use 1 to indicate your most preferred layout and 4 to indicate your least preferred. An example response would be: Layout 1=3, Layout 2=4, Layout 3=2, Layout 4=1
    \begin{enumerate}
        \item [i]Layout 1: Text followed by chart(s) \rule{2cm}{0.4pt}
        \item [ii]Layout 2: Chart(s) followed by text \rule{2cm}{0.4pt} 
        \item [iii]Layout 3: Chart(s) within the text \rule{2cm}{0.4pt}
        \item [iv]Layout 4: Text only \rule{2cm}{0.4pt}
    \end{enumerate}

    \item Please explain your ranking in Question 8. \\ \\
    \noindent\rule{7cm}{0.4pt}\\ 
    \noindent\rule{7cm}{0.4pt} \\ 
    
\end{enumerate}

\noindent
\textbf{Section 2 of 2: Decision-Making with Reports} \\

\begin{enumerate}
    \setcounter{enumi}{9}
    
    \item Of the decision types listed below, please estimate the percentage of decision(s) you typically make for each. An example answer would be "Strategic= 70\%, Tactical= 10\%, Operational=10\%, Other=10\%." Note, percentages should add up to 100\%.
    \begin{enumerate}
        \item [i]Strategic \rule{2cm}{0.4pt}
        \item [ii]Tactical \rule{2cm}{0.4pt} 
        \item [iii]Operational \rule{2cm}{0.4pt}
        \item [iv]Other (Please specify) \rule{2cm}{0.4pt}
    \end{enumerate} 

\noindent
1) \textbf{Strategic decisions} are long-term choices that set the course of an organization. E.g., Should we change existing policies to address an emerging problem? \\
2) \textbf{Tactical decisions} are choices about how things will get done. E.g., What specific changes should we make to the policy? \\
3) \textbf{Operational decisions} refer to choices that employees make each day to make the organization run. E.g., What immediate action should I take to ensure the maximum effective impact on a problem? 


    \item Which layout format do you think is most ideal for Strategic decisions?
        \begin{enumerate}
        \item [i]Layout 1: Text followed by chart(s) \rule{2cm}{0.4pt}
        \item [ii]Layout 2: Chart(s) followed by text \rule{2cm}{0.4pt} 
        \item [iii]Layout 3: Chart(s) within the text \rule{2cm}{0.4pt}
        \item [iv]Layout 4: Text only \rule{2cm}{0.4pt}
        \item [iv]Other \rule{2cm}{0.4pt}
    \end{enumerate}

    \item Which layout format do you think is most ideal for Tactical decisions?
        \begin{enumerate}
        \item [i]Layout 1: Text followed by chart(s) \rule{2cm}{0.4pt}
        \item [ii]Layout 2: Chart(s) followed by text \rule{2cm}{0.4pt} 
        \item [iii]Layout 3: Chart(s) within the text \rule{2cm}{0.4pt}
        \item [iv]Layout 4: Text only \rule{2cm}{0.4pt}
        \item [iv]Other \rule{2cm}{0.4pt}
    \end{enumerate}

    \item Which layout format do you think is most ideal for Operational decisions?
        \begin{enumerate}
        \item [i]Layout 1: Text followed by chart(s) \rule{2cm}{0.4pt}
        \item [ii]Layout 2: Chart(s) followed by text \rule{2cm}{0.4pt} 
        \item [iii]Layout 3: Chart(s) within the text \rule{2cm}{0.4pt}
        \item [iv]Layout 4: Text only \rule{2cm}{0.4pt}
        \item [iv]Other \rule{2cm}{0.4pt}
    \end{enumerate}

        \item Which layout format do you think is most ideal for Other decisions?
        \begin{enumerate}
        \item [i]Layout 1: Text followed by chart(s) \rule{2cm}{0.4pt}
        \item [ii]Layout 2: Chart(s) followed by text \rule{2cm}{0.4pt} 
        \item [iii]Layout 3: Chart(s) within the text \rule{2cm}{0.4pt}
        \item [iv]Layout 4: Text only \rule{2cm}{0.4pt}
        \item [iv]Other \rule{2cm}{0.4pt}
    \end{enumerate}
    
    \item Describe the steps you take from reading the report to making a decision. What do you pay attention to? \\ \\
    \noindent\rule{7cm}{0.4pt}\\ 
    \noindent\rule{7cm}{0.4pt} \\ 

    \item Do your reports have different levels of time sensitivity?
    \begin{itemize}
        \item [$\ocircle$] Yes
        \item [$\ocircle$] No 
    \end{itemize}

    \item If yes, describe how a time-sensitive report might differ from a non-time-sensitive report. \\ \\
    \noindent\rule{7cm}{0.4pt}\\ 
    \noindent\rule{7cm}{0.4pt} \\

\end{enumerate}

\bibliographystyle{abbrv}

\bibliography{main}

\end{multicols}
\end{document}